\newcounter{myctr}
\def\myitem{\refstepcounter{myctr}\bibfont\noindent\ifnum\themyctr>9\else\phantom{0}\fi\hangindent17pt\themyctr.\enskip}

\documentclass{ws-ijqi}

\usepackage{braket}

\begin{document}

\markboth{Takuya Machida}
{Localization Model of 2-State QWs}

\catchline{}{}{}{}{}

\title{LIMIT THEOREMS FOR A LOCALIZATION\\MODEL OF 2-STATE QUANTUM WALKS}

\author{TAKUYA MACHIDA}

\address{Meiji Institute for Advanced Study of Mathematical Sciences,\\
Meiji University, 1-1-1 Higashimita, Tamaku,\\
Kawasaki 214-8571, Japan\\
bunchin@meiji.ac.jp}

\maketitle

\begin{history}
\received{12 October 2010}

\end{history}

\begin{abstract}
We consider 2-state quantum walks (QWs) on the line, which are defined by two matrices.
One of the matrices operates the walk at only half-time.
In the usual QWs, localization does not occur at all.
However, our walk can be localized around the origin.
In this paper, we present two limit theorems, that is, one is a stationary distribution and the other is a convergence theorem in distribution.
\end{abstract}



\section{Introduction}

The 2-state quantum walk (QW) on the line $\mathbb{Z}=\left\{\,0,\pm 1,\pm 2,\ldots\,\right\}$ has been intensively studied, and the limit theorems are obtained.\cite{aharonov,ambainis_2001,kempe,kendon,konno_2008_2}
For example, the limit distribution of the usual walks was calculated.\cite{konno_2002_1,konno_2005_1}
In the present paper, we consider a localization model of 2-state QWs.
The motivation is the analysis of the time-inhomogeneous QWs.
Our walks are determined by two matrices, one of which operates the walk at only half-time.
The walk can be considered as one of the time-dependent models, for which there are some results.\cite{banuls,romanelli_2009_1,machida_2010_1}
Particularly, Ref. \refcite{banuls} and \refcite{romanelli_2009_1} discuss localization.
We present the two limit theorems that show the localization of the probability distribution.
One is calculation of the limit value for the probability $P(X_t=x)$ which walker is at position $x\in\mathbb{Z}$ starting from the origin, where $t\in\left\{0,1,2,\ldots\right\}$ is time, and the other is the convergence in distribution.
In the usual walks, localization can not occurs.
However, if we change the matrix at only half-time, then we find that the localization occurs from the results in this paper.
The localization of QWs, which can be applied to quantum search,\cite{reitzner_2009} is often investigated.\cite{chisaki,inui1,inui3,konno_2009_2,wojcik,watabe2008limit}
If $\limsup_{t\rightarrow\infty}P(X_t=x)>0$ for a position $x$, we call that the localization occurs.
Therefore, our result insists that the localization occurs for any initial state.
Moreover, we obtain the convergence in distribution of $X_t/t$ as $t\rightarrow\infty$.
This limit distribution is described by both $\delta$-function and a density function.
For 3-state Grover walk, similar limit theorems were shown.\cite{inui1}
The limit distribution of a 4-state walk corresponding to the 2-state walk with memory was also computed.\cite{machida_2010_2,gettrick}
The present paper is organized as follows.
In Section 2, we define our walk.
We present the limit theorems as our main result in Section 3.
Section 4 is devoted to the proofs of the theorems.
By using the Fourier analysis, we obtain the limit distribution.
Summary is given in the final section.


\section{Definition of a localization model of 2-state QWs}

In this section, we define a localization model of 2-state QWs on the line.
Let $\ket{x}$ ($x\in\mathbb{Z}$) be an infinite components vector which denotes the position of the walker.
Here,  $x$-th component of $\ket{x}$ is 1 and the other is 0.
Let $\ket{\psi_{t}(x)} \in \mathbb{C}^2$ be the amplitude of the walker at position $x$ at time $t$.
The walk at time $t$ is expressed by
\begin{equation}
 \ket{\Psi_t}=\sum_{x\in\mathbb{Z}}\ket{x}\otimes\ket{\psi_{t}(x)}.
\end{equation}
The time evolution of our walk is depicted with the following two unitary matrices:
\begin{align}
  U=&\left[\begin{array}{cc}
     \cos\theta &\sin\theta \\\sin\theta &-\cos\theta
	  \end{array}\right]
  =\left[\begin{array}{cc}
    c & s \\ s &-c
	 \end{array}\right],\\
  H=&\left[\begin{array}{cc}
     \cos\theta_1 &\sin\theta_1 \\\sin\theta_1 &-\cos\theta_1
	  \end{array}\right]
  =\left[\begin{array}{cc}
   c_1 & s_1 \\ s_1 &-c_1
	 \end{array}\right],
\end{align}
where $c=\cos\theta,s=\sin\theta\,(\theta\in [0,2\pi),\,\theta\neq 0,\frac{\pi}{2},\pi,\frac{3\pi}{2})$ and $c_1=\cos\theta_1,s_1=\sin\theta_1\,(\theta_1\in [0,2\pi))$.
Moreover, we introduce four matrices:
\begin{equation}
 P=\left[\begin{array}{cc}
    c & s\\ 0&0
	 \end{array}\right],\,
 Q=\left[\begin{array}{cc}
    0&0\\ s &-c
	 \end{array}\right],\,
 P_1=\left[\begin{array}{cc}
	    c_1 & s_1\\ 0&0
		 \end{array}\right],\,
 Q_1=\left[\begin{array}{cc}
	    0&0\\ s_1 &-c_1
		 \end{array}\right].
\end{equation}
Then, the evolution is determined by
\begin{equation}
 \ket{\psi_{t+1}(x)}
 =\left\{\begin{array}{ll}
   P\ket{\psi_t(x+1)}+Q\ket{\psi_t(x-1)}&(t\neq \tau)\\[3mm]
    P_1\ket{\psi_t(x+1)}+Q_1\ket{\psi_t(x-1)}&(t=\tau)
	 \end{array}\right. ,\label{eq:te}
\end{equation}
where $\tau\in\left\{1,2,\ldots\right\}$.
Note that $P+Q=U$ and $P_1+Q_1=H$.
The probability that the quantum walker $X_t$ is at position $x$ at time $t$, $P(X_t=x)$, is defined by
\begin{equation}
 P(X_t=x)=\braket{\psi_t(x)|\psi_t(x)}.\label{eq:prob}
\end{equation}
In our main results, we focus on the probability distribution at time $2\tau+1, 2\tau+2$.
So, time $\tau$ is called half-time in our walk.

\begin{figure}[h]
 \begin{center}
 \begin{minipage}{60mm}
  \begin{center}
   \includegraphics[scale=0.45]{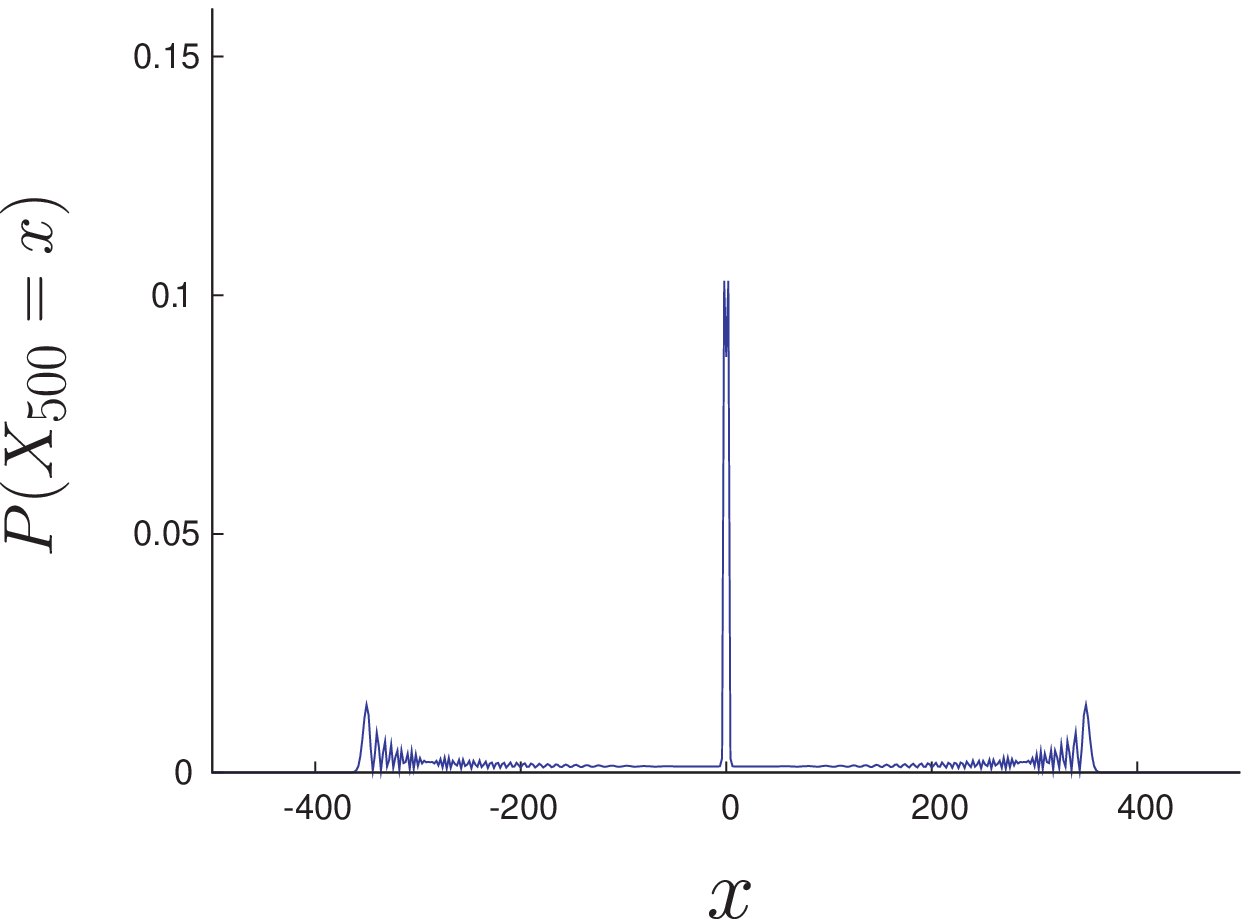}\\
   {(a) $\ket{\psi_0(0)}={}^T[1/\sqrt{2}\,,i/\sqrt{2}\,]$}
  \end{center}
 \end{minipage}
 \begin{minipage}{60mm}
  \begin{center}
   \includegraphics[scale=0.45]{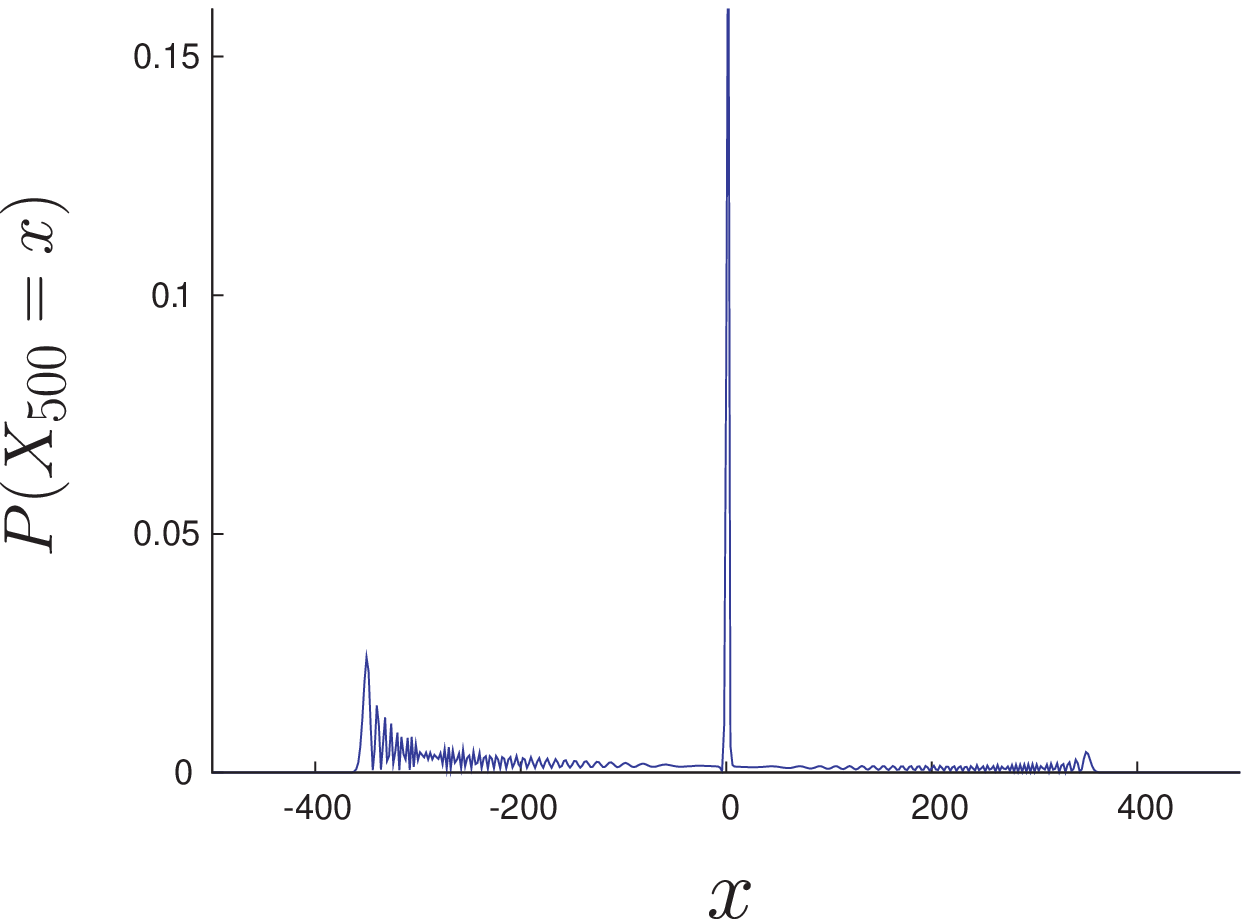}\\
   {(b) $\ket{\psi_0(0)}={}^T[1,0]$}
  \end{center}
 \end{minipage}
 \vspace{5mm}
 \caption{The probability distributions at time $t=500$ as $\tau=249$ and $\theta=\pi/4,\theta_1=0$.}
  \label{fig:distribution}
 \end{center}
\end{figure}

\begin{figure}[h]
 \begin{center}
 \begin{minipage}{60mm}
  \begin{center}
   \includegraphics[scale=0.6]{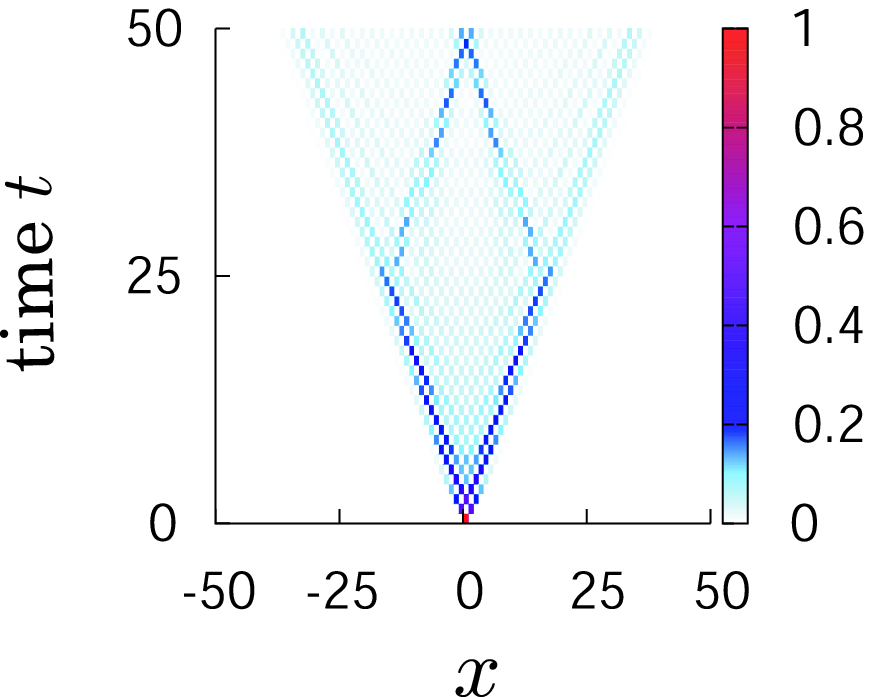}\\
   {(a) $\ket{\psi_0(0)}={}^T[1/\sqrt{2}\,,i/\sqrt{2}\,]$}
  \end{center}
 \end{minipage}
 \begin{minipage}{60mm}
  \begin{center}
   \includegraphics[scale=0.6]{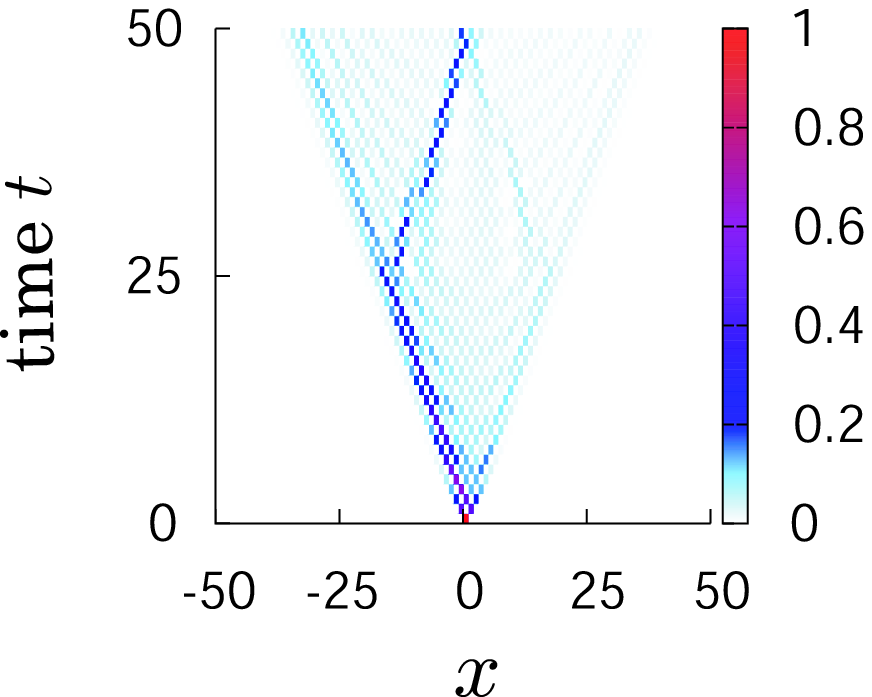}\\
   {(b) $\ket{\psi_0(0)}={}^T[1,0]$}
  \end{center}
 \end{minipage}
 \vspace{5mm}
 \caption{The evolution of the probability distributions for time $t$ by density plot as $\tau=24$ and $\theta=\pi/4,\theta_1=0$.}
  \label{fig:dis_time}
 \end{center}
\end{figure}

The Fourier transform $\ket{\hat{\Psi}_{t}(k)}\,(k\in\left[-\pi,\pi\right))$ of $\ket{\psi_t(x)}$ is given by
\begin{equation}
 \ket{\hat{\Psi}_{t}(k)}=\sum_{x\in\mathbb{Z}} e^{-ikx}\ket{\psi_t(x)}.\label{eq:ft}
\end{equation}
By the inverse Fourier transform, we have
\begin{equation}
 \ket{\psi_t(x)}=\int_{-\pi}^{\pi}e^{ikx}\ket{\hat\Psi_{t}(k)}\,\frac{dk}{2\pi}.
\end{equation}
From (\ref{eq:te}) and (\ref{eq:ft}), the time evolution of $\ket{\hat{\Psi}_{t}(k)}$ becomes
\begin{equation}
 \ket{\hat{\Psi}_{t+1}(k)}=
  \left\{\begin{array}{ll}
   \hat U(k)\ket{\hat{\Psi}_{t}(k)}&(t\neq\tau)\\[2mm]
	  \hat H(k)\ket{\hat{\Psi}_{t}(k)}&(t=\tau)
	 \end{array}\right.,\label{eq:timeevo}
\end{equation}
where $\hat U(k)=R(k)U,\,\hat{H}(k)=R(k)H$ and
$
 R(k)=\left[\begin{array}{cc}
       e^{ik}&0\\
	     0&e^{-ik}
	    \end{array}\right]
$.
Particularly, we see that
\begin{align}
 \ket{\hat{\Psi}_{2\tau+1}(k)}=&\hat U(k)^\tau \hat H(k)\hat U(k)^\tau\ket{\hat\Psi_{0}(k)},\\
 \ket{\hat{\Psi}_{2\tau+2}(k)}=&\hat U(k)^{\tau+1} \hat H(k)\hat U(k)^\tau\ket{\hat\Psi_{0}(k)}.\label{eq:ft_even}
\end{align}
In the present paper, we take the initial state as
\begin{equation}
 \ket{\psi_0(x)}=\left\{\begin{array}{ll}
		 \!{}^T[\,\alpha, \,\beta\,]& (x=0)\\[2mm]
			\!{}^T[\,0,\,0\,]& (x\neq 0)
		       \end{array}\right.,
\end{equation}
where $|\alpha|^2+|\beta|^2=1$ and $T$ is the transposed operator.
We should note that $\ket{\hat\Psi_{0}(k)}=\ket{\psi_0(0)}$.
Figures \ref{fig:distribution} and \ref{fig:dis_time} depict probability distributions of the walk under the condition $\theta=\pi/4,\,\theta_1=0$.
In Figure \ref{fig:distribution} (a), the probabilities at the position $x=0,\pm 2$ are higher than other positions.
Figure \ref{fig:dis_time} shows the time evolution of probability distribution as $\tau=24$.
The walk evolves with the matrix $H$ at only time $\tau$.
So, the walk is a usual QW till time $\tau$.

\begin{figure}[h]
 \begin{center}
 \begin{minipage}{60mm}
  \begin{center}
   \includegraphics[scale=0.45]{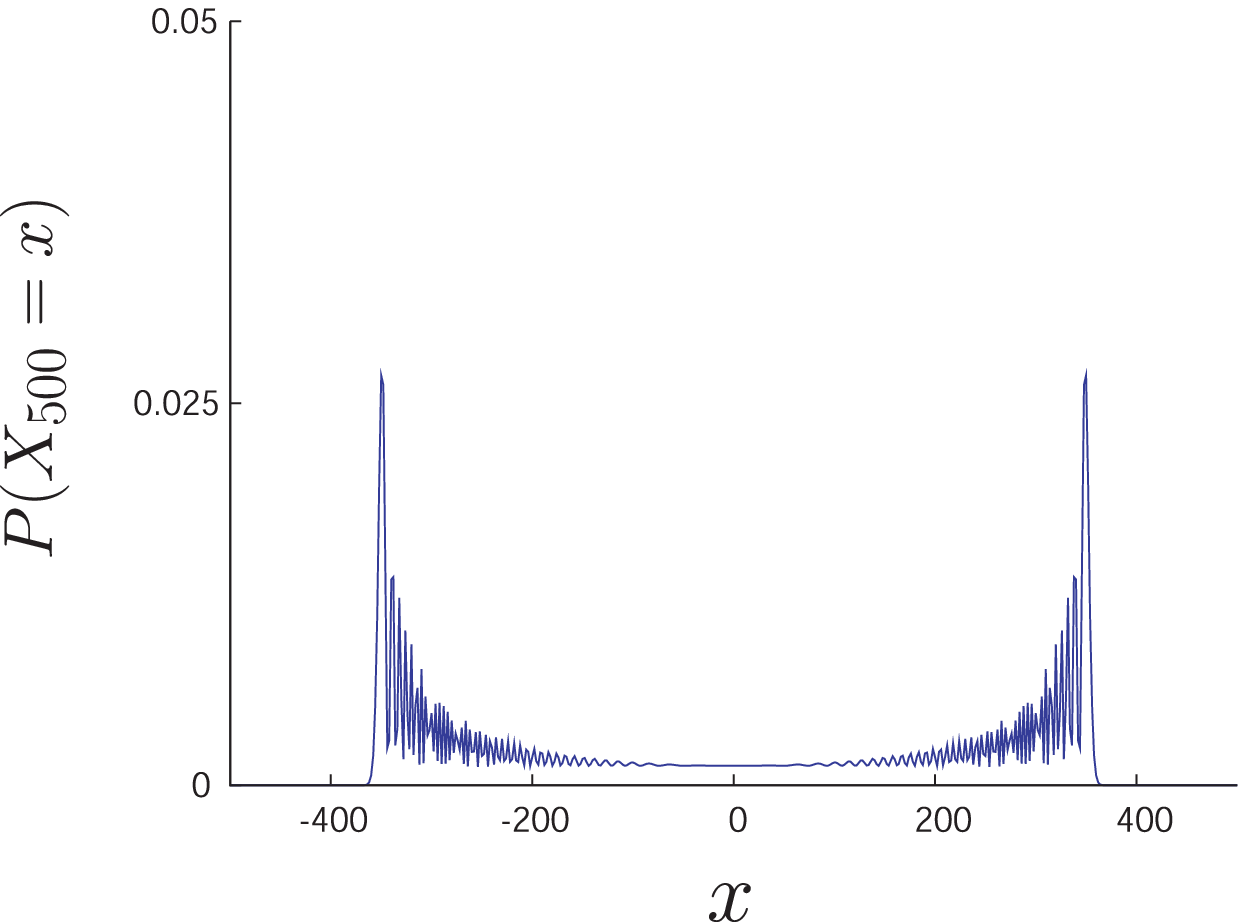}\\
   {(a) $\ket{\psi_0(0)}={}^T[1/\sqrt{2}\,,i/\sqrt{2}\,]$}
  \end{center}
 \end{minipage}
 \begin{minipage}{60mm}
  \begin{center}
   \includegraphics[scale=0.45]{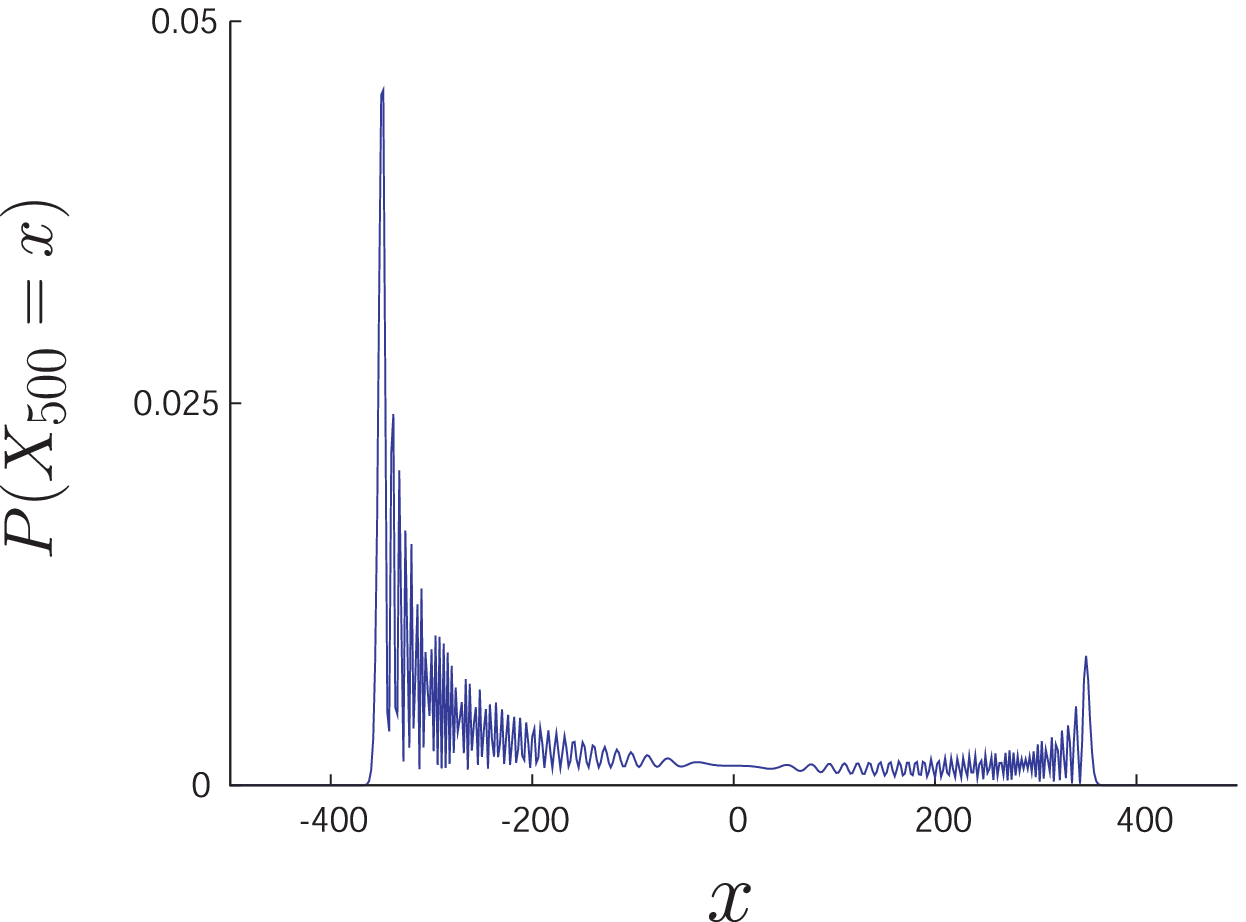}\\
   {(b) $\ket{\psi_0(0)}={}^T[1,0]$}
  \end{center}
 \end{minipage}
 \vspace{5mm}
 \caption{The probability distributions at time $t=500$ with $\theta=\theta_1=\pi/4$.}
  \label{fig:distribution_usualQW}
 \end{center}
\end{figure}

\begin{figure}[h]
 \begin{center}
 \begin{minipage}{60mm}
  \begin{center}
   \includegraphics[scale=0.6]{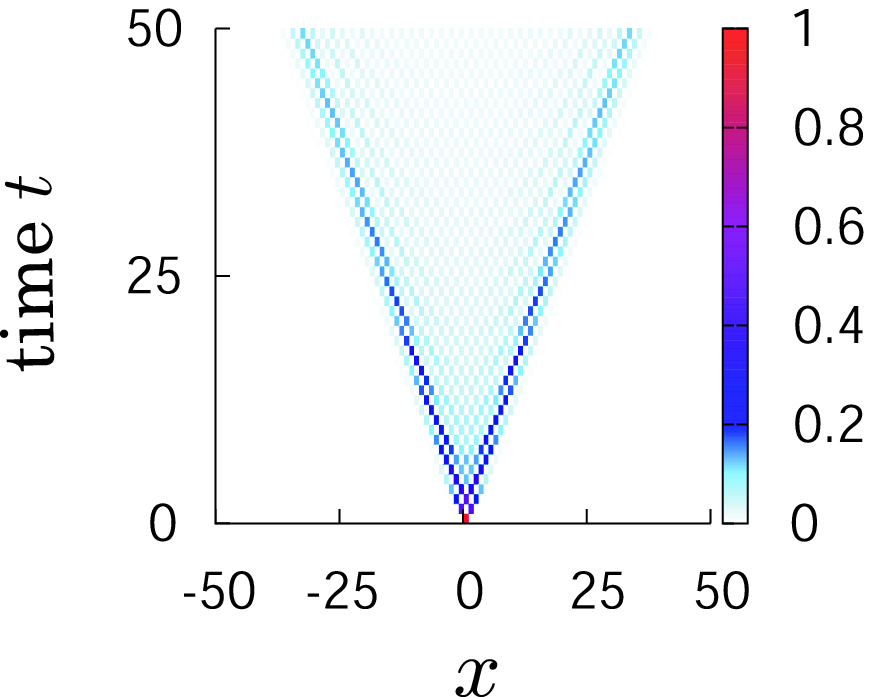}\\
   {(a) $\ket{\psi_0(0)}={}^T[1/\sqrt{2}\,,i/\sqrt{2}\,]$}
  \end{center}
 \end{minipage}
 \begin{minipage}{60mm}
  \begin{center}
   \includegraphics[scale=0.6]{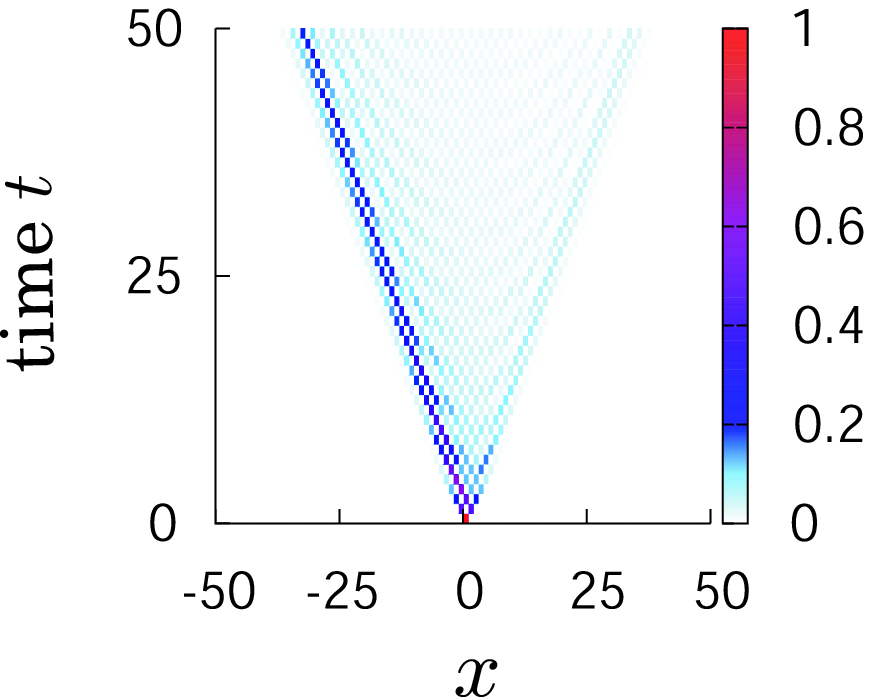}\\
   {(b) $\ket{\psi_0(0)}={}^T[1,0]$}
  \end{center}
 \end{minipage}
 \vspace{5mm}
 \caption{The evolution of the probability distributions for time $t$ by density plot with $\theta=\theta_1=\pi/4$.}
  \label{fig:dis_time_usualQW}
 \end{center}
\end{figure}

 In comparison with both Figures \ref{fig:distribution} and \ref{fig:dis_time}, we show the probability distributions of a usual QW with $\theta=\theta_1=\pi/4$ in Figures \ref{fig:distribution_usualQW} and \ref{fig:dis_time_usualQW}.

\clearpage


\section{Limit theorems for the walk}

In this section, we show our main results.
For our localization model of 2-state QWs, we obtain the two following limit theorems.

\begin{theorem}

\noindent {\it (i) For odd time $2\tau+1$,}
\begin{equation}
 \lim_{\tau\rightarrow\infty}P(X_{2\tau+1}=x)=\left\{\begin{array}{cl}
					       \frac{(c_1s-s_1c)^2(1-|s|)^2}{c^6}K_1(-1;\beta,\alpha)&(x=-1),\\[5mm]
						\frac{(c_1s-s_1c)^2(1-|s|)^2}{c^6}K_1(1;\alpha,\beta)&(x=1),\\[5mm]
						\frac{2(c_1s-s_1c)^2s}{c^4(1-|s|)}\left(\frac{1-|s|}{c}\right)^{2|x|}K_2(x; \beta,\alpha)&(x=-3,-5,-7,\ldots),\\[5mm]
					 	\frac{2(c_1s-s_1c)^2s}{c^4(1-|s|)}\left(\frac{1-|s|}{c}\right)^{2|x|}K_2(x; \alpha,\beta)&(x=3,5,7,\ldots),\\[5mm]
						0&(x=0,\pm 2,\pm 4,\ldots),
					     \end{array}\right.
\end{equation}
{\it where}
\begin{align}
 K_1(x; \alpha,\beta)=&c^2|\alpha|^2+2s^2(1-|s|)|\beta|^2+sign(x)cs(1-|s|)(\alpha\overline{\beta}+\overline{\alpha}\beta),\\
 K_2(x; \alpha,\beta)=&c^2s|\alpha|^2+s(1-|s|)^2|\beta|^2-sign(x) c|s|(1-|s|)(\alpha\overline{\beta}+\overline{\alpha}\beta),
\end{align}
{\it and $sign(x)$ means a sign of $x$.}\\

\noindent {\it (ii) For even time $2\tau+2$,}
\begin{equation}
 \lim_{\tau\rightarrow\infty}P(X_{2\tau+2}=x)=\left\{\begin{array}{cl}
					       \frac{(c_1s-s_1c)^2s^2(1-|s|)^2}{c^4}&(x=0),\\[5mm]
						\frac{(c_1s-s_1c)^2(1-|s|)^2}{c^8}K_3(-2;\beta,\alpha)&(x=-2),\\[5mm]
						\frac{(c_1s-s_1c)^2(1-|s|)^2}{c^8}K_3(2;\alpha,\beta)&(x=2),\\[5mm]
						\frac{2(c_1s-s_1c)^2|s|}{c^4}\left(\frac{1-|s|}{c}\right)^{2|x|}K_4(x; \beta,\alpha)&(x=-4,-6,-8,\ldots),\\[5mm]
						\frac{2(c_1s-s_1c)^2|s|}{c^4}\left(\frac{1-|s|}{c}\right)^{2|x|}K_4(x; \alpha,\beta)&(x=4,6,8,\ldots),\\[5mm]
						0&(x=\pm 1,\pm 3,\ldots),
					     \end{array}\right.
\end{equation}
{\it where}
\begin{align}
 K_3(x;\alpha,\beta)=&c^2\left\{1-s^2|s|(2-|s|)\right\}|\alpha|^2+2s^2(1-|s|)^3|\beta|^2\nonumber\\
 &+sign(x)cs(1+s^2)(1-|s|)^2(\alpha\overline{\beta}+\overline{\alpha}\beta),\\[2mm]
 K_4(x;\alpha,\beta)=&s^2(|\alpha|^2-|\beta|^2)-sign(x)cs(\alpha\overline{\beta}+\overline{\alpha}\beta)+|s|.
\end{align}
\end{theorem}

\begin{figure}[h]
 \begin{center}
 \begin{minipage}{40mm}
  \begin{center}
   \includegraphics[scale=0.3]{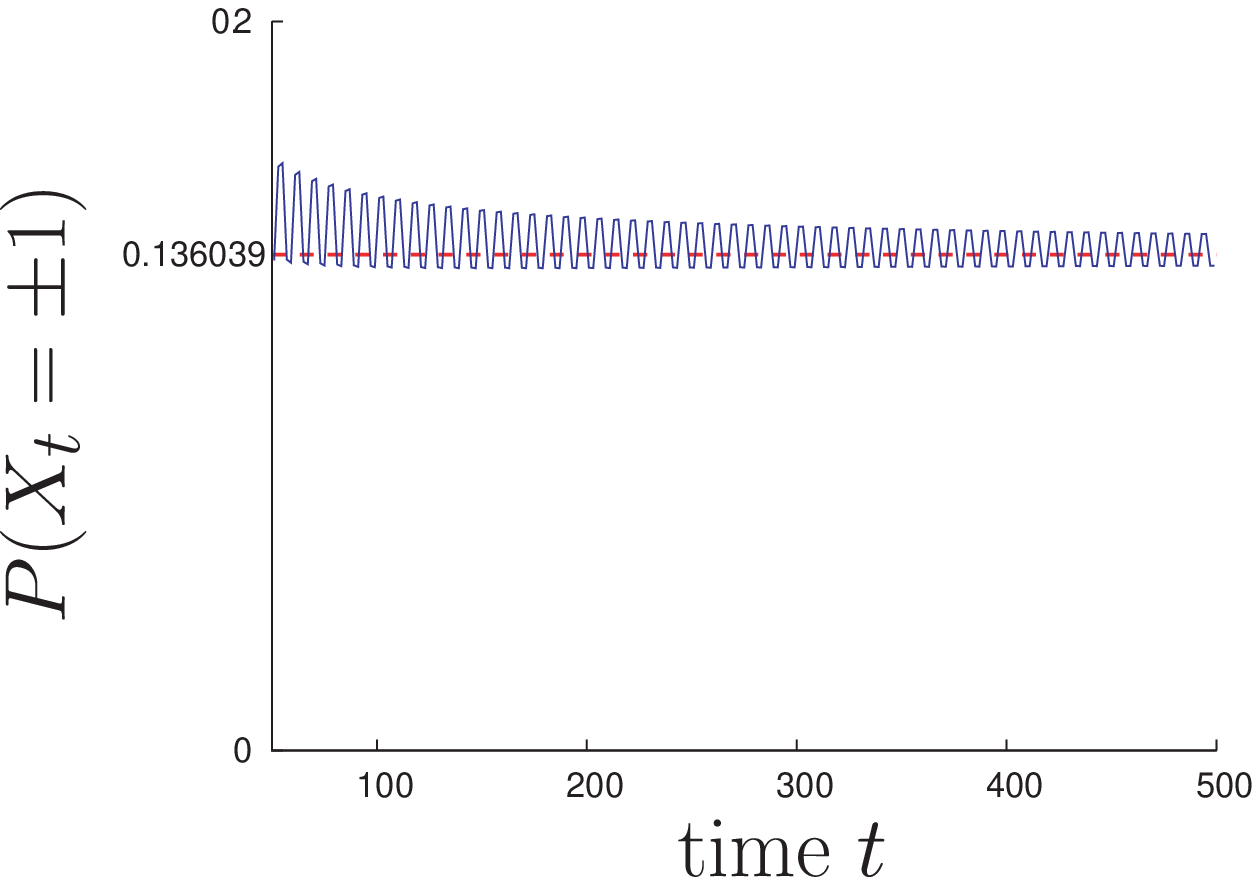}\\
   {(a)}
  \end{center}
 \end{minipage}\hspace{1cm}
 \begin{minipage}{40mm}
  \begin{center}
   \includegraphics[scale=0.3]{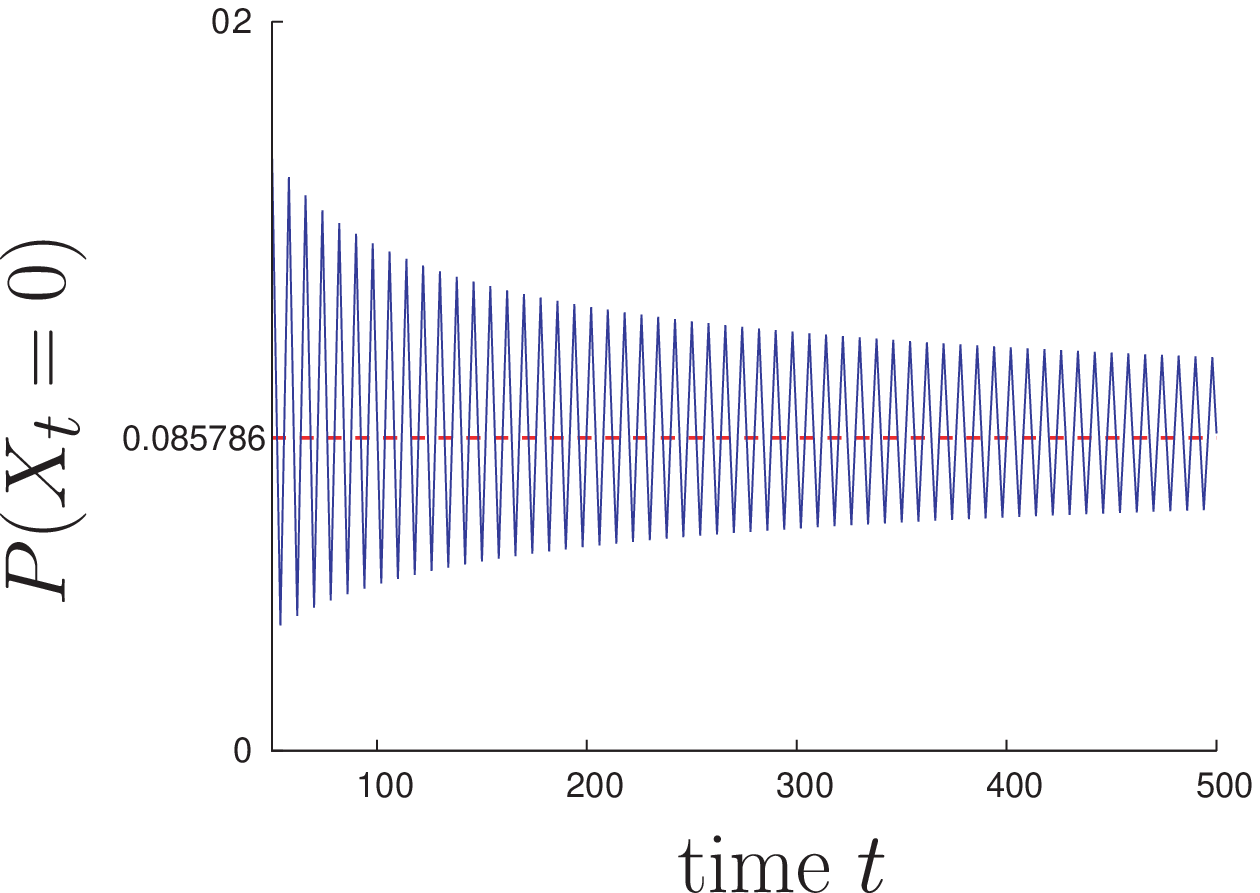}\\
   {(b)}
  \end{center}
 \end{minipage}

  \begin{minipage}{40mm}
  \begin{center}
   \includegraphics[scale=0.3]{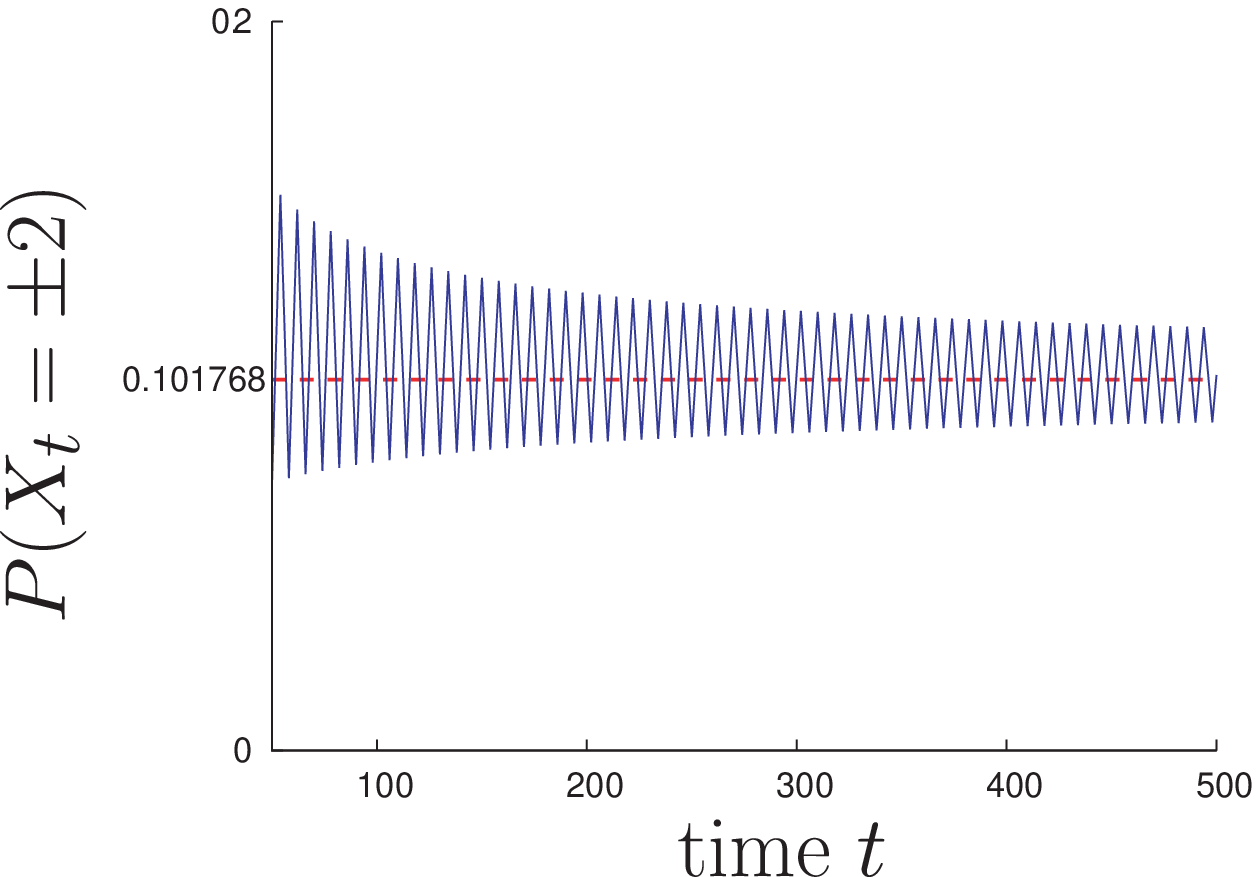}\\
   {(c)}
  \end{center}
  \end{minipage}
  \vspace{5mm}
 \caption{The behaviors of the probability $P(X_t=0, \pm 1,\pm 2)$ as $\theta=\pi/4,\theta_1=0$.}
  \label{fig:p0}
 \end{center}
\end{figure}

\vspace{3mm}
\noindent
As a simple example of the case when the localization occurs in our walk, if $\theta=\pi/4,\theta_1=0$ and $\ket{\psi_0(0)}={}^T[1/\sqrt{2}\,,i/\sqrt{2}\,]$, then we have
\begin{align}
 \lim_{\tau\rightarrow\infty}P(X_{2\tau+1}=\pm 1)=&\frac{13-9\sqrt{2}}{2}=0.136039\cdots,\label{eq:prob_012_1}\\
 \lim_{\tau\rightarrow\infty}P(X_{2\tau+2}=0)=&\frac{3-2\sqrt{2}}{2}=0.0857864\cdots,\label{eq:prob_012_0}\\
 \lim_{\tau\rightarrow\infty}P(X_{2\tau+2}=\pm 2)=&\frac{139-98\sqrt{2}}{4}=0.101768\cdots.\label{eq:prob_012_2}
\end{align}
Figure \ref{fig:p0} corresponds to the behavior of each probability in (\ref{eq:prob_012_1}), (\ref{eq:prob_012_0}) and (\ref{eq:prob_012_2}).


Next, we present the theorem of the convergence in distribution for $X_t/t$, where $t\in\left\{2\tau+1,2\tau+2\right\}$.
Some similar results corresponding to Theorem 2 were shown for a 3-state walk or a 4-state walk.\cite{inui1,machida_2010_2}
Moreover, localization of multi-state walks was computed.\cite{inui3}
The limit distribution of the usual 2-state walk does not have the delta measure.\cite{konno_2002_1,konno_2005_1}
However, we find that the limit distribution of the 2-state walk defined in this paper has a delta-measure from Theorem 2.\\

\begin{theorem}
{\it For our localization model of 2-state QWs, we have}
\begin{equation}
 \lim_{t\rightarrow\infty} P(X_t/t\leq x)=\int_{-\infty}^{x}f(x)\,dx,
\end{equation}
{\it where}
\begin{align}
 f(x)=&\Delta\delta_0(x)+\frac{|s|}{\pi(1-x^2)\sqrt{c^2-x^2}}\left[1-\left\{|\alpha|^2-|\beta|^2+\frac{(\alpha\overline{\beta}+\overline{\alpha}\beta)s}{c}\right\}x\right]\nonumber\\
 &\times \frac{a_2x^4+a_1x^2+a_0}{c^2(1-x^2)}\,I_{(-|c|,|c|)}(x),
\end{align}
{\it and $\delta_0(x)$ denotes the delta-measure at the origin and $I_{A}(x)=1$ if $x\in A$, $I_{A}(x)=0$ if $x\notin A$.
The values $\Delta,a_0,a_1,a_2$ are independent on initial state $\ket{\psi_0(0)}$ as follows:}
\begin{align}
 \Delta=&\frac{(c_1s-s_1c)^2}{1+|s|},\\
 a_0=&c^2,\\
 a_1=&2s_1c(c_1s-s_1c)-c_1^2,\\
 a_2=&(c_1s-s_1c)^2.
\end{align}
{\it Particularly, if $\theta_1=\theta$, then we obtain the density function of the usual walk, that is,}
\begin{align}
 f(x)=&\frac{|s|}{\pi(1-x^2)\sqrt{c^2-x^2}}\left[1-\left\{|\alpha|^2-|\beta|^2+\frac{(\alpha\overline{\beta}+\overline{\alpha}\beta)s}{c}\right\}x\right]\,I_{(-|c|,|c|)}(x).
\end{align}
\end{theorem}

\begin{figure}[h]
 \begin{center}
 \begin{minipage}{60mm}
  \begin{center}
   \includegraphics[scale=0.4]{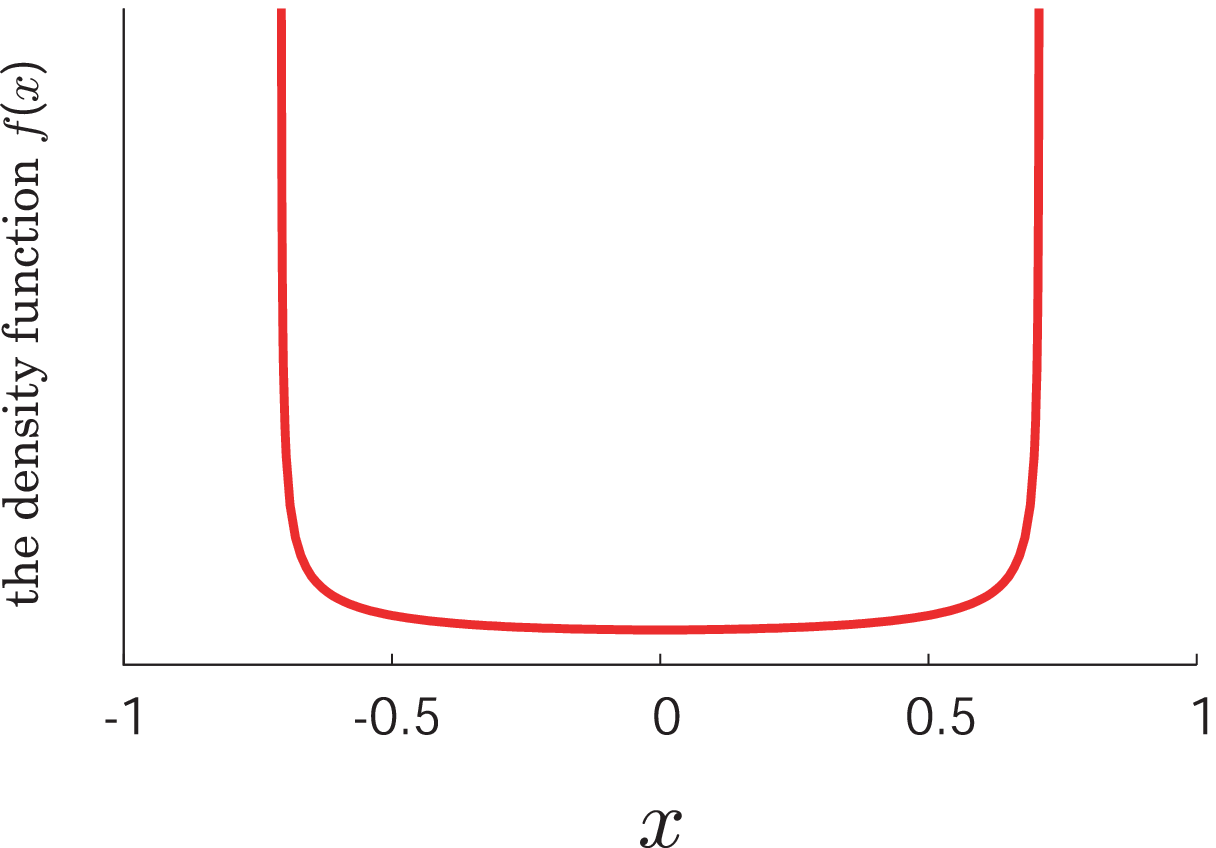}\\
   {(a) $\ket{\psi_0(0)}={}^T[1/\sqrt{2}\,,i/\sqrt{2}\,]$}
  \end{center}
 \end{minipage}
 \begin{minipage}{60mm}
  \begin{center}
   \includegraphics[scale=0.4]{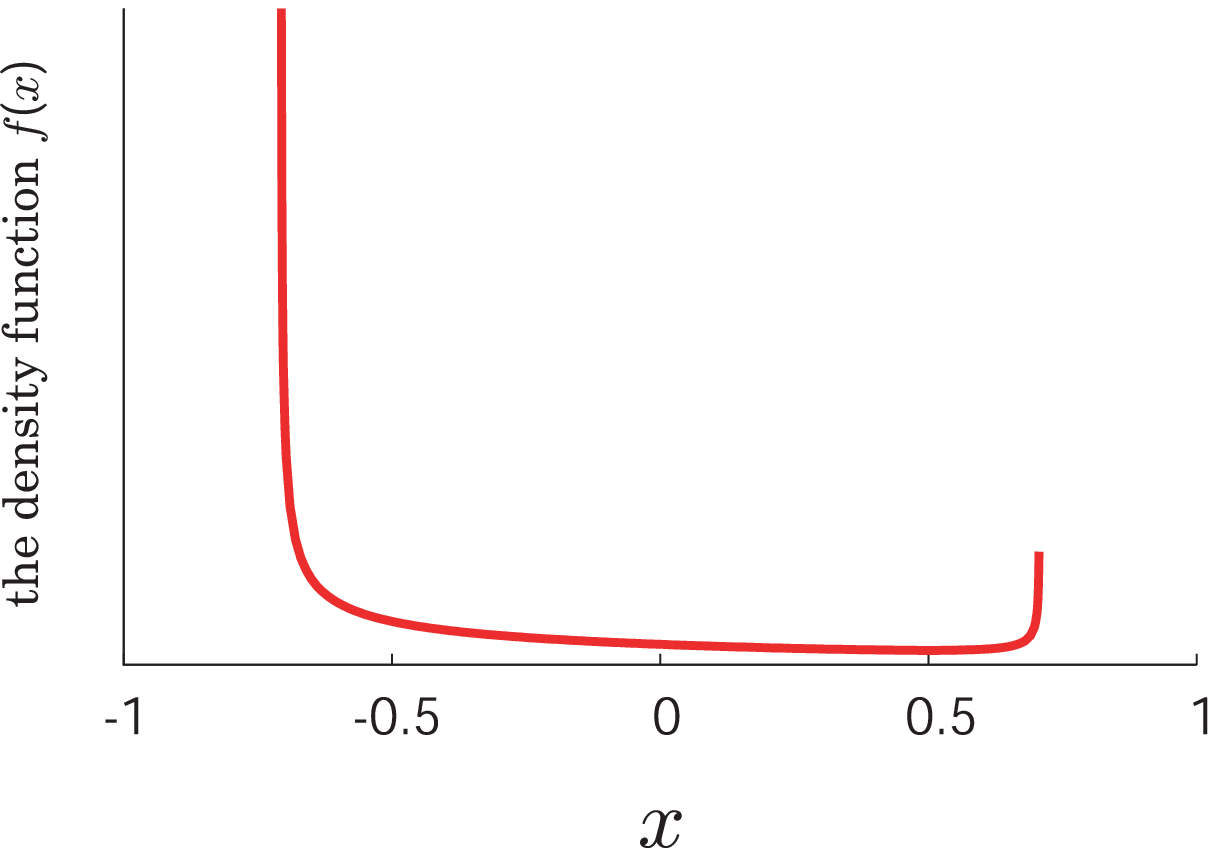}\\
   {(b) $\ket{\psi_0(0)}={}^T[1,0]$}
  \end{center}
 \end{minipage}
  \vspace{5mm}
 \caption{The limit density function as $\theta=\pi/4,\theta_1=0$.}
  \label{fig:density}
 \end{center}
\end{figure}

Figure \ref{fig:density} shows the density function $f(x)$ with $\theta=\pi/4,\theta_1=0$.
We have $\Delta=\frac{1}{2+\sqrt{2}}$ in Figure \ref{fig:density}.

We should note that the following relation between Theorems 1 and 2:
\begin{equation}
 \sum_{x\in\mathbb{Z}}\lim_{t\rightarrow\infty}P(X_t=x)=\Delta.
\end{equation}
Therefore, $\lim_{t\to\infty}P(X_t=x)$ is not a probability measure.

\section{Proofs of theorems}

In this section, we will prove Theorems 1 and 2 in Section 3. 
Our approach is based on the Fourier analysis.\cite{grimmett}


\subsection{Proof of Theorem 1}

At first, the eigenvalues $\lambda_j(k)\,(j=1,2)$ of $\hat U(k)$ can be computed as
\begin{equation}
 \lambda_1(k)=\sqrt{1-c^2\sin^2 k}+ic\sin k,\,\lambda_2(k)=-\sqrt{1-c^2\sin^2 k}+ic\sin k.
\end{equation}
The normalized eigenvector $\ket{v_j(k)}$ corresponding to $\lambda_j(k)$ is
\begin{equation}
  \ket{v_j(k)}=\sqrt{\frac{\sqrt{1-c^2\sin^2 k}\pm c\cos k}{2s^2\sqrt{1-c^2\sin^2 k}}}
   \left[\begin{array}{c}
    se^{ik}\\ \pm\sqrt{1-c^2\sin^2 k}-c\cos k
	 \end{array}\right].
\end{equation}
Therefore, the Fourier transform $\ket{\hat\Psi_{0}(k)}$ is expressed by $\ket{v_j (k)}$ as follows:
\begin{equation}
 \ket{\hat\Psi_{0}(k)}=\sum_{j=1}^2\braket{v_j(k)|\hat\Psi_{0}(k)}\ket{v_j(k)}.\label{eq:psi_0_kiyosato}
\end{equation}
In the proof, we focus on even time $2\tau+2$.
From (\ref{eq:ft_even}) and (\ref{eq:psi_0_kiyosato}), the Fourier transform at time $2\tau+2$ is given by
\begin{align}
 \ket{\hat\Psi_{2\tau+2}(k)}=&\hat U(k)^{\tau+1} \hat H(k)\hat U(k)^\tau\ket{\hat\Psi_{0}(k)}\nonumber\\
 =&\hat U(k)^{\tau+1} \hat H(k)\sum_{j_1=1}^2\lambda_{j_1}^\tau(k)\braket{v_{j_1}(k)|\hat\Psi_0(k)}\ket{v_{j_1}(k)}\nonumber\\
 =&\sum_{j_1=1}^2\lambda_{j_1}^\tau(k)\braket{v_{j_1}(k)|\hat\Psi_0(k)}\left(\hat U(k)^{\tau+1} \hat H(k)\ket{v_{j_1}(k)}\right).
\end{align}
Moreover, rewriting $\hat U(k)^{\tau+1}\hat H(k)\ket{v_{j_1}(k)}$ as
\begin{align}
 \hat U(k)^{\tau+1}\hat H(k)\ket{v_{j_1}(k)}=&\hat U(k)^{\tau+1}\sum_{j_2=1}^2\braket{v_{j_2}(k)|\hat H(k)|v_{j_1}(k)}\ket{v_{j_2}(k)}\nonumber\\
 =&\sum_{j_2=1}^2\lambda_{j_2}^{\tau+1}(k)\braket{v_{j_2}(k)|\hat H(k)|v_{j_1}(k)}\ket{v_{j_2}(k)},
\end{align}
we obtain
\begin{align}
 \ket{\hat\Psi_{2\tau+2}(k)}
 =&\sum_{j_1=1}^2\sum_{j_2=1}^2\lambda_{j_1}(k)^{\tau}\lambda_{j_2}(k)^{\tau+1}\braket{v_{j_1}(k)|\hat\Psi_0(k)}\nonumber\\
 &\qquad\qquad\times\braket{v_{j_2}(k)|\hat H(k)|v_{j_1}(k)}\ket{v_{j_2}(k)}.\label{eq:psi_2t+2}
\end{align}
We should note $\lambda_1(k)\lambda_2(k)=-1$.
Calculating the inverse Fourier transform, we have
\begin{align}
 \ket{\psi_{2\tau+2}(x)}=&
\sum_{j_1=1}^2\sum_{j_2=1}^2\,\int_{-\pi}^{\pi}\lambda_{j_1}(k)^{\tau}\lambda_{j_2}(k)^{\tau+1}\braket{v_{j_1}(k)|\hat\Psi_0(k)}\nonumber\\
 &\qquad\qquad\times\braket{v_{j_2}(k)|\hat H(k)|v_{j_1}(k)}\ket{v_{j_2}(k)}e^{ikx}\,\frac{dk}{2\pi}.
\end{align}
By using the Riemann-Lebesgue lemma,\cite{inui1} we see
\begin{align}
 \ket{\psi_{2\tau+2}(x)}\sim&\quad(-1)^\tau\int_{-\pi}^{\pi}\lambda_1(k)\braket{v_2(k)|\hat\Psi_0(k)}\braket{v_1(k)|\hat H(k)|v_2(k)}\ket{v_1(k)}e^{ikx}\,\frac{dk}{2\pi}\nonumber\\
 &+(-1)^\tau\int_{-\pi}^{\pi}\lambda_2(k)\braket{v_1(k)|\hat\Psi_0(k)}\braket{v_2(k)|\hat H(k)|v_1(k)}\ket{v_2(k)}e^{ikx}\,\frac{dk}{2\pi},\label{eq:psi_t_sim}
\end{align}
where $g(\tau)\sim h(\tau)$ denotes $\lim_{\tau\rightarrow\infty}g(\tau)/h(\tau)=1$.
From (\ref{eq:psi_t_sim}), we get
\begin{align}
 \ket{\psi_{2\tau+2}(x)}\,\sim&\,\left\{
				   \begin{array}{cl}
				    \frac{(-1)^\tau(c_1s-s_1c)|s|(1-|s|)}{c^2}\left[\begin{array}{c}
								   -\beta\\\alpha
									\end{array}\right]&(x=0),\\[5mm]
				    \frac{(-1)^\tau(c_1s-s_1c)}{c^2}
				     \left[\begin{array}{c}
				      -csI_2\alpha-(1-|s|)\beta\\
					    I_2\left\{|s|(1-|s|)\alpha+cs\beta\right\}
					   \end{array}\right]&(x=-2),\\[5mm]
				    \frac{(-1)^\tau(c_1s-s_1c)}{c^2}
				     \left[\begin{array}{c}
				      I_2\left\{cs\alpha-|s|(1-|s|)\beta\right\}\\
					    (1-|s|)\alpha-csI_2\beta
					   \end{array}\right]&(x=2),\\[5mm]
				    \frac{(-1)^\tau(c_1s-s_1c)I_x}{c^2}
				     \left[\begin{array}{c}
				      \pm cs\alpha-|s|(1\mp |s|)\beta\\
					    |s|(1\pm |s|)\alpha\mp cs\beta
					   \end{array}\right]&(x=\pm 4,\pm 6,\ldots),\\[5mm]
				     \left[\begin{array}{c}
				      0\\0
					   \end{array}\right]&(x=\pm 1,\pm 3,\ldots),
				   \end{array}\right.\label{eq:psi_even}
\end{align}
where $I_x=\left\{\frac{i(1-|s|)}{|c|}\right\}^{|x|}$.
Similarly we can compute $\ket{\psi_{2\tau+1}(x)}$ as follows:
\begin{align}
 \ket{\psi_{2\tau+1}(x)}\,\sim&\,\left\{
				   \begin{array}{cl}
				    \frac{(-1)^\tau(c_1s-s_1c)(1-|s|)}{c^3}
				     \left[\begin{array}{c}
				      c(s\alpha-c\beta)\\
					-|s|(1-|s|)\alpha-cs\beta
					   \end{array}\right]&(x=-1),\\[5mm]
				    \frac{(-1)^\tau(c_1s-s_1c)(1-|s|)}{c^3}
				     \left[\begin{array}{c}
				      cs\alpha-|s|(1-|s|)\beta\\
					-c(c\alpha+s\beta)
					   \end{array}\right]&(x=1),\\[5mm]
				     (-1)^\tau J_x
				     \left[\begin{array}{c}
				      -c^2\left\{s(1-|s|)\alpha+c|s|\beta\right\}\\
					    (1-|s|)\left\{c|s|(1-|s|)\alpha+c^2s\beta\right\}
					   \end{array}\right]&(x=-3,-5,\ldots),\\[5mm]
				     (-1)^\tau J_x
				     \left[\begin{array}{c}
				      (1-|s|)\left\{-c^2s\alpha+c|s|(1-|s|)\beta\right\}\\
				       -c^2\left\{c|s|\alpha-s(1-|s|)\beta\right\}
					   \end{array}\right]&(x=3,5,\ldots),\\[5mm]
				     \left[\begin{array}{c}
				      0\\0
					   \end{array}\right]&(x=0,\pm 2,\pm 4,\ldots),
				   \end{array}\right.\label{eq:psi_odd}
\end{align}
where $J_x=\frac{i(c_1s-s_1c)I_x}{c^2|c|(1-|s|)}$.
From (\ref{eq:prob}), (\ref{eq:psi_even}) and (\ref{eq:psi_odd}), the proof is completed.
\begin{flushright}
$\Box$ 
\end{flushright}


\subsection{Proof of Theorem 2}

We calculate the characteristic function $E(e^{izX_{t}/t})$ as $t\rightarrow\infty$, where $E(X)$ denotes the expected value of $X$.
At first, (\ref{eq:psi_2t+2}) can be written as
\begin{align}
 \ket{\hat\Psi_{2\tau+2}(k)}=&\hat U(k)^{\tau+1}\hat H(k)\hat U(k)^\tau\ket{\hat\Psi_{0}(k)}\nonumber\\
 =&\lambda_1(k)^{2\tau+2}A_1(k)\ket{v_1(k)}+\lambda_2(k)^{2\tau+2}A_2(k)\ket{v_2(k)}\nonumber\\
 &-\lambda_3(k)^{2\tau+2}A_3(k)\ket{v_1(k)}-\lambda_4(k)^{2\tau+2}A_4(k)\ket{v_2(k)},
\end{align}
where $\lambda_3(k)=\lambda_4(k)=i$ and
\begin{align}
 A_1(k)=&\overline\lambda_1(k)\braket{v_1(k)|\hat\Psi_0(k)}\braket{v_1(k)|\hat H(k)|v_1(k)},\\
 A_2(k)=&\overline\lambda_2(k)\braket{v_2(k)|\hat\Psi_0(k)}\braket{v_2(k)|\hat H(k)|v_2(k)},\\
 A_3(k)=&\lambda_1(k)\braket{v_2(k)|\hat\Psi_0(k)}\braket{v_1(k)|\hat H(k)|v_2(k)},\\
 A_4(k)=&\lambda_2(k)\braket{v_1(k)|\hat\Psi_0(k)}\braket{v_2(k)|\hat H(k)|v_1(k)}.
\end{align}
Substituting $2\tau+2=t$, we can calculate the {\it r}-th moment of $X_t$ as follows:
\begin{align}
 E((X_{t})^r)=&\sum_{x\in \mathbb{Z}}x^r P(X_{t}=x)\nonumber\\
=&\int_{-\pi}^{\pi}\bra{\hat\Psi_{t}(k)}\left(D^r\ket{\hat\Psi_{t}(k)}\right)\,\frac{dk}{2\pi}\nonumber\\
 =&(t)_r\int_{-\pi}^{\pi}\Biggl\{\,\sum_{j=1}^4 h_j(k)^r|A_j|^2-\lambda_1(k)^t\overline{\lambda_3}(k)^t h_1(k)^r A_1(k)\overline{A_3}(k)\nonumber\\
 &-\overline{\lambda_1}(k)^t\lambda_3(k)^t h_3(k)^r \overline{A_1}(k)A_3(k)-\lambda_2(k)^t\overline{\lambda_4}(k)^t h_2(k)^r A_2(k)\overline{A_4}(k)\nonumber\\
 &-\overline{\lambda_2}(k)^t\lambda_4(k)^t h_4(k)^r \overline{A_2}(k)A_4(k)\Biggr\}\,\frac{dk}{2\pi}\,+O(t^{r-1}),
\end{align}
where $h_j(k)=D\lambda_j(k)/\lambda_j(k)$, $D=i(d/dk)$ and $(t)_r=t(t-1)\times\cdots\times(t-r+1)$.
By using the Riemann-Lebesgue lemma, we have
\begin{align}
 \lim_{t\rightarrow\infty}E((X_{t}/t)^r)
  =&\int_{-\pi}^{\pi}\sum_{j=1}^4 h_j^r(k)|A_j(k)|^2\,\frac{dk}{2\pi}\nonumber\\
  =&0^r\Delta+\int_{-\pi}^{\pi}\sum_{j=1}^2 h_j^r(k)|A_j(k)|^2\,\frac{dk}{2\pi},
\end{align}
where
\begin{equation}
 \Delta=\int_{-\pi}^{\pi}\sum_{j=3}^4 |A_j(k)|^2\,\frac{dk}{2\pi}=\frac{(c_1s-s_1c)^2}{1+|s|}.
\end{equation}
Therefore, we obtain
\begin{align}
 \lim_{t\rightarrow\infty}E((X_{t}/t)^r)=&0^r\Delta
 +\int_{-\infty}^{\infty} x^r\frac{|s|}{\pi(1-x^2)\sqrt{c^2-x^2}}\left[1-\left\{|\alpha|^2-|\beta|^2+\frac{(\alpha\overline{\beta}+\overline{\alpha}\beta)s}{c}\right\}x\right]\nonumber\\
 &\times\frac{a_2x^4+a_1x^2+a_0}{c^2(1-x^2)}\,I_{(-|c|,|c|)}(x)\,dx\nonumber\\
=&\int_{-\infty}^{\infty} x^r f(x)\,dx,\label{eq:r-th_mom}
\end{align}
where
\begin{align}
 a_0=&c^2,\\
 a_1=&2s_1c(c_1s-s_1c)-c_1^2,\\
 a_2=&(c_1s-s_1c)^2.
\end{align}
We should remark
\begin{equation}
 \int_{-\infty}^{\infty} g(x)\delta_0(x)\,dx=g(0).
\end{equation}
\clearpage

\noindent By (\ref{eq:r-th_mom}), we can compute the characteristic function $E(e^{izX_{t}/t})$ as $t\rightarrow\infty$.
Thus the proof of Theorem 2 is completed.
\begin{flushright}
$\Box$ 
\end{flushright}

\section{Summary}

In the final section, we conclude and discuss the probability distribution of our walks.
In the usual 2-state walk defined by the matrix $U$, the localization does not occur at all.
However, if another matrix $H(\neq U)$ operates the walk at only half-time, then localization occurs.
In Theorem 1, the behavior of the probability $P(X_t=x)$ was calculated as $t\to\infty$.
Moreover, we found that the limit distribution of $X_t/t$ had both a delta measure and a density function from Theorem 2.
The interesting problem is calculation of the limit distribution for the walk which the matrix $H$ operates more than twice.

\section*{Acknowledgments}
\noindent
The author is grateful to Norio Konno for useful comment and also to Joe Yuichiro Wakano and the Meiji University Global COE Program ``Formation and Development of Mathematical Sciences Based on Modeling and Analysis'' for the support.


\end{document}